\documentclass[twocolumn,preprintnumbers,amssymb,amsmath,superscriptaddress]{revtex4}
\usepackage[dvips]{graphicx}
\usepackage{dcolumn}
\usepackage{bm}
\usepackage{epsfig,amsmath,amssymb,verbatim,mathrsfs,array,layout,textcomp,amssymb,latexsym}
\input epsf.tex
\newcommand{\beq}{\begin{eqnarray}}
\newcommand{\eeq}{\end{eqnarray}}

\def\ltap{\ \raise.3ex\hbox{$<$\kern-.75em\lower1ex\hbox{$\sim$}}\ }
\def\gtap{\ \raise.3ex\hbox{$>$\kern-.75em\lower1ex\hbox{$\sim$}}\ }

\def\ie{{\it i.e.}}

\def\be{\begin{equation}}
\def\ee{\end{equation}}
\def\bea{\begin{eqnarray}}
\def\eea{\end{eqnarray}}

\newcommand{\gev}{{\rm GeV}}
\newcommand{\tev}{{\rm TeV}}

\def\mysection#1{{{\bf #1}.~}}

\begin{document}

\title{
Probing Dark Matter Dynamics via Earthborn Neutrinos at IceCube
}
\author{C\'edric Delaunay}
\affiliation{Department of Particle Physics, Weizmann Institute of Science, Rehovot 76100, Israel}
\author{Patrick J. Fox}
\affiliation{Theoretical Physics Department, FNAL
 Batavia, IL 60510,USA}
\author{Gilad Perez}
\affiliation{Department of Particle Physics, Weizmann Institute of Science, Rehovot 76100, Israel}
\preprint{\scriptsize FERMILAB-PUB-08-568-T, WIS/23/08-DEC-DPP\vspace*{.1cm}\\}
\vskip .05in

\begin{abstract}
\vskip .05in
Recent results from PAMELA and ATIC hint that $\mathcal{O}(\tev)$ dark matter (DM) is annihilating, in our galactic neighborhood, predominantly to leptons.  The annihilation rate is much larger now than during freeze-out, one possible explanation of this is a low-velocity 
enhancement of the annihilation cross section.  In a model independent fashion, we show that in this case the rate of neutrino emission from the Earth, due to DM annihilation, may be greatly enhanced while the rate from the Sun is unaltered.  
There is potential for IceCube to see these earthborn neutrinos while the same parameter space will be soon covered by direct detection experiments. Combining these near-future data will allow extraction of valuable information about the DM sector dynamics. 
\end{abstract}

\maketitle
Approximately 20\% of the matter-energy budget of the universe is due to Dark Matter (DM).  The favored candidate for the DM particle is a thermal relic with annihilation cross section $\langle \sigma v \rangle \approx 3\times 10^{-26} \mathrm{cm}^3$s$^{-1}$, a weakly interacting DM (WIMP).  Many experiments are underway to probe the DM, either directly through its interactions with Standard Model (SM) particles or indirectly through its annihilations to SM particles.  Recently several indirect detection experiments have reported results which may be interpreted as due to DM annihilations, 
although they could also have an astrophysical origin~\cite{Hooper:2008kg}.

The ATIC experiment has reported an excess of electron-positron flux around $300-800\,\gev$~\cite{:2008zz}. In addition PAMELA~\cite{Adriani:2008zr} is seeing an increase of the positron fraction around energies of $10-80\,\gev$ 
and no corresponding excess in the antiproton fraction~\cite{Adriani:2008zq}.
  Taken together
these suggest that ${\cal O}(1\rm\,TeV)$ DM, annihilating preferentially to leptons, is being observed~\cite{Cirelli:2007xd,Cirelli:2008pk,ArkaniHamed:2008qn}.  However, the annihilation cross section required to explain the excesses is substantially larger than $3\times 10^{-26} \mathrm{cm}^3$s$^{-1}$ 
~\cite{Cirelli:2008jk,Cholis:2008hb}.  The enhancement may be due to a boost factor, a nearby clump of DM or a low-velocity Sommerfeld effect (see also~\cite{Feldman:2008xs,Ibe:2008ye} for an alternative explanation).

These results are exciting and surprising, not only 
are we possibly observing WIMP DM but maybe also a non-trivial DM sector, whose dynamics seems to imply an epoch-dependent annihilation cross section.
Looking at the photon spectrum~\cite{Collaboration:2008aa,Hinton:2004eu,deBoer:2005tm,Elsaesser:2004xa,Cecchi:2008zz} and additional cosmic ray (CR) experiments~\cite{Boezio:2001ac,Barwick:1997ig,Aguilar:2007yf} will test this emerging paradigm~\cite{ArkaniHamed:2008qn} and whether these excesses are actually due to DM.
However, the photon and CR flux depends on the DM halo profile and the propagation model.
We demonstrate here that the same DM sector dynamics may
induce dramatic changes in the neutrino flux from the Earth which give a very different probe of the same microscopic phenomena. For a possible signal from galactic neutrinos see~\cite{Liu:2008ci,Hisano:2008ah}.

Dense bodies such as the Sun and the Earth gravitationally capture DM particles in their core, resulting in a DM density significantly higher than in the galactic halo.  
They eventually start to annihilate into SM particles, among which neutrinos can escape and travel to Earth-based detectors. The resulting flux depends on
the capture and annihilation cross sections, unless the DM has already reached equilibrium which leads to a maximal flux, exclusively controlled by the capture rate.
The effects which enhance the annihilation rate would, generically, not
affect the capture rate. For instance, an ultra light particle with sizable coupling to the nuclei is required to Sommerfeld enhance the capture rate which is probably in conflict
with various precision data. 
While the capture rate remains unaltered, a larger annihilation cross section will shorten the typical time for the DM to reach equilibrium.  Our key observation is that since it is very probable that  the Earth has not yet
reached equilibrium for a relic annihilation cross section~\cite{Gould:1987ir, Lundberg:2004dn}, 
this effect would
yield order of magnitudes enhancement in the neutrino flux from the core of the Earth at IceCube~\cite{Achterberg:2006md}. Moreover 
this flux will also be correlated with the DM direct search experiments~\cite{Ahmed:2008eu,Angle:2007uj}. 
A combination of these
two data sets yields fairly clean information about the microscopic nature of the dark sector dynamics.

\mysection{Neutrinos via DM annihilation}
The competition between capture and annihilation of the DM leads to a present day DM annihilation rate~\cite{Jungman:1995df} 
\be
\label{eqn:gamma}
\Gamma =\frac{1}{2}A N^2 = \frac{C}{2}\tanh^2 \left(t_\oplus\sqrt{CA}\right)~,
\ee
where $t_\oplus \simeq 4.5\times 10^9$ yrs is the age of the Earth, $A=\langle \sigma v \rangle/V_{eff}$, $V_{eff}= 5.7\times 10^{22} \mathrm{cm}^3 \left(\tev /m_\chi\right)^{3/2}$ is the effective volume of the core of the Earth \cite{Gould:1987ir} and $C$ is the capture rate.  For the Earth the capture rate is dominated by the spin independent (SI) elastic scattering (see~\cite{TuckerSmith:2001hy} for the inelastic case) of the DM off various elements in the Earth~\cite{Jungman:1995df},
\be
C_\oplus \simeq 1.7
\times 10^5\mbox{s}^{-1}\frac{\rho^\chi_{0.3}}{\left(v^{\chi}_{270}\right)^3}\left(\frac{\tev}{m_\chi}\right)^2\sum_i f_i \left(\frac{\sigma^{N_i}_{\rm SI}}{10^{-6}\mathrm{pb}}\right)~,
\ee
where the the sum is over the elements O, Si, Mg, S, Fe and Ni, only 3\% of the mass of the Earth is neglected. The DM mass is denoted $m_\chi$, $\rho^\chi_{0.3}$ and $v^\chi_{270}$ are the DM energy density and velocity in the halo in units of 0.3~$\,\gev/$cm$^3$ and 270~km$/$s respectively, while the factor $f_i$ accounts for the mass fraction and distribution profile of the element $i$~\cite{Jungman:1995df}, whose cross section with DM is denoted $\sigma_{\rm SI}^{N_i}$.  
Direct detection experiments probe SI cross section of DM off proton, $\sigma^p_{\rm SI}$.  To better than 1$\%$, $\sigma^N_{\rm SI}\approx N^4\sigma^p_{\rm SI}$ for any nucleus of mass number $N$.
Hence, 
\be\label{eqn:caprate}
C_\oplus \simeq 9.6
\times 10^{11}\mbox{s}^{-1}\frac{\rho^\chi_{0.3}}{\left(v^{\chi}_{270}\right)^3}\left(\frac{\mbox{TeV}}{m_\chi}\right)^2\left(\frac{\sigma^p_{\rm SI}}{10^{-6}\mbox{pb}}\right).
\ee
The maximum rate of DM annihilation occurs after equilibrium is reached and is entirely determined by the capture rate, $\Gamma_{\rm eq}=C/2$. 
For times shorter than the equilibrium time $t_{\rm eq}=1/\sqrt{CA}$ the abundance is grows linearly with time and the annihilation rate is 
$
\Gamma_{\rm neq} \sim \frac{1}{2}A C^2t^2.
$
With a typical thermal relic annihilation cross section, $A_{\rm r}\simeq 5.3\times 10^{-49}{\rm s}^{-1}\left(m_\chi/\tev\right)^{3/2}$, the Earth is 
far from equilibrium ($t_\oplus\ll t_{\rm eq}$) and not a good source of DM-neutrinos. However if the observed electrons/positrons excesses are due to a low-velocity enhancement, $R$, the annihilation cross section can be far larger than that of the early universe, $A_\oplus=R A_{\rm r}$, bringing the Earth towards equilibrium today.  
The maximal enhancement in the rate 
is
$
\Gamma_{\rm eq}/\Gamma_{\rm neq}\sim (A_{\rm r} C_\oplus t_\oplus^2)^{-1}
$
which can be several orders of magnitude and is obtained for $R\gtrsim (A_{\rm r} C_\oplus t_\oplus^2)^{-1}$.
The escape velocity at the center of the Earth is approximately $15\,\mathrm{km\ s}^{-1}$ whilst DM in the halo has a Maxwell-Boltzmann distribution with $v_0=270\,\mathrm{km\ s}^{-1}$.  The Sommerfeld enhancement grows as $\sim 1/v$ although this growth saturates at very low velocities~\cite{Cirelli:2007xd}, a further increase beyond $v=270\,\mathrm{km\ s}^{-1}$ may yield more non-trivial information
about the DM sector.  Thus, the enhancement may in fact be even larger than that for DM in the halo.  
It will be useful to define the critical capture rate for the Earth:
\be\label{CcriticEarth}
C^{\rm c}_\oplus=1/A_{\rm r} t_\oplus^{2}\simeq 9.93 \times 10^{13}\mathrm{s}^{-1}\left(\frac{\tev}{m_\chi}\right)^{3/2}~,
\ee
above which the Earth would already have reached equilibrium 
and boosting the annihilation cross section will not result in an 
enhanced neutrino flux.
Direct searches experiment such as CDMSII put an upper bound~\cite{Ahmed:2008eu} on the SI elastic scattering cross section of $3.5\times 10^{-7}$pb for $m_\chi=1\, \tev$.
Thus, $C_\oplus\lesssim 10^{-2}C^{\rm c}_\oplus$ and the Earth is probably still far from equilibrium.

The capture rate (\ref{eqn:caprate}) is derived assuming that the DM velocity distribution as encountered by the earth is Gaussian.  It is possible that in the solar system it differs from Gaussian~\cite{Lundberg:2004dn,Peter:2008sy}, particularly at the low velocities necessary for capture in the Earth and Sun.  The DM abundance may also differ significantly from the galactic halo density~(see {\it e.g.} \cite{Damour:1998vg, Xu:2008ep, Lundberg:2004dn, Peter:2008sy, Vogelsberger:2008qb} and Refs. therein).  Both direct and indirect detection experiments probe the same nuclear scattering cross section but they are sensitive to the different parts of the velocity distribution, high and low velocity respectively. 
Assuming a Gaussian distribution allows observations from direct and  indirect experiments to be straightforwardly correlated.
Furthermore, a future signal at direct detection experiments would directly probe velocity distributions (through differential energy information) 
of the DM particles~\cite{Angle:2007uj,Brink:2005ej,Aprile:2002ef} at ranges of roughly 40-150\,km\,s$^{-1}$. 
Of particular importance are the Xe based experiments which have the lowest threshold, down to approximately three times the earth escape velocity~\cite{Aprile:2002ef,LUX}.

\begin{figure}[tp]
\begin{center}
\includegraphics[width=200pt]{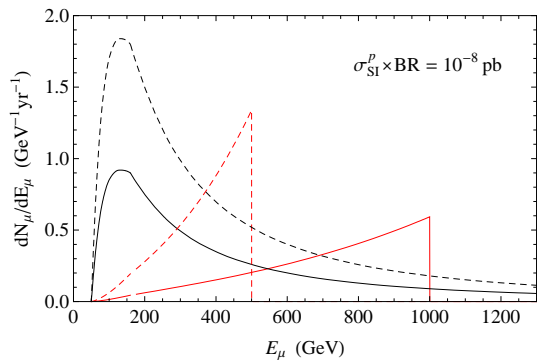}
\caption{Muon rates from primary neutrinos (red) and atmospheric backgrounds (black) at IceCube.  
All plots show results for a 1 TeV (solid) and 500 GeV (dashed) DM, corresponding angular cuts have been placed on the background.  The signal assumes the Earth has reached equilibrium.
}\label{fig:primaryrates}
\end{center}
\end{figure}

\mysection{Annihilation into primary neutrinos}
The muon flux at the surface of the Earth is given by:
\bea
\frac{d\Phi_\mu^{\rm P}}{dE_\mu}&= & \int_{E_\mu}^\infty dE_\nu\frac{d\Phi_\nu}{dE_\nu}\left[\frac{d\sigma_\nu^p(E_\nu,E_\mu)}{dE_\mu}\rho_p+(p\rightarrow n)
\right]\nonumber \\ 
&& \times~R_\mu(E_\mu)+(\nu  \rightarrow \bar{\nu})~,
\eea
with $\rho_{p,n}$ the number density of protons and neutrons in the medium, respectively $5/9 N_A$cm$^{-3}$ and $4/9 N_A$cm$^{-3}$ for ice, where $N_A\simeq 6\times 10^{23}$ is Avogadro's number. $d\sigma^{p,n}_\nu/dE_\mu$ are the weak scattering cross sections of (anti-) neutrinos on nucleons 
\be
\frac{d\sigma_\nu^{p,n}}{dE_\mu}= \frac{2m_p G_F^2}{\pi}\left(a^{p,n}_\nu+b^{p,n}_\nu\frac{E_\mu^2}{E_{\nu}^2}\right)~,
\ee
where 
$a^{n,p}_\nu=0.25,0.15$, $b^{n,p}_\nu=0.06,0.04$ and $a_{\bar{\nu}}^{n,p}=b_\nu^{p,n}$, $b_{\bar{\nu}}^{n,p}=a_\nu^{p,n}$~\cite{Barger:2007xf}.
The distance $R_\mu(E_\mu)$, 
the muon range, defines the distance traveled by a muon until its energy drops below the energy threshold $E_{\rm th}$ of the detector, due to losses 
in the medium.   Approximately,
$R_\mu(E_\mu)=\frac{1}{\rho\beta}\log\left[\frac{\alpha+\beta E_\mu}{\alpha+\beta E_{\rm th}}\right],$
with $\rho$ the density of the medium ($\simeq 1$g\,cm$^{-3}$ for ice) 
and $\alpha\simeq 2.0$ MeV\,cm$^2$g$^{-1}$ and $\beta\simeq 4.2\times 10^{-6}$cm$^2$g$^{-1}$ for ice. At IceCube, the energy threshold is about 50 GeV and for $E_\mu\sim$ TeV, the typical muon range is a few kilometers, which is longer than the detector typical size~\footnote{The higher density rock bed below IceCube stops a significant number of muons due to a larger number of interaction~\cite{Dutta:2000hh}. 
However, the production rate of muons will be enhanced for the same reason and both effects cancel to leading order.}. 
Since the DM is almost at rest, the muon neutrino flux at the surface of the Earth is monochromatic,
$
d\Phi_\nu/dE_\nu=\delta(E_\nu-m_\chi) B_{\bar{\nu}\nu}\Gamma/4\pi R_\oplus^2~,
$ 
with $B_{\bar{\nu}\nu}$ the branching ratio of DM annihilating to neutrino pair and $R_\oplus\simeq 6.4\times 10^3$km, the Earth radius. The resulting muon flux is:
\bea
\frac{d\Phi_\mu}{dE_\mu}&=& \frac{B_{\bar{\nu}\nu}\Gamma}{4\pi R_\oplus^2} \left[\frac{d\sigma_\nu^p(m_\chi,E_\mu)}{dE_\mu}\rho_p+(p\rightarrow n)
\right]\nonumber \\ 
&& \times ~R_\mu(E_\mu)\Theta(m_\chi-E_\mu)+(\nu  \rightarrow \bar{\nu}).
\eea 
Combining this with the effective area~\cite{GonzalezGarcia:2005xw} of the detector $A_{eff}(E_\mu)$ gives the event rate in the detector, \ie\ $dN/dE_\mu = A_{eff}(E_\mu) d\Phi_\mu/dE_\mu$.  This is shown for DM masses of $500\,\gev$ and $1\, \tev$ in Figure~\ref{fig:primaryrates}, along with the background rate (discussed below) from atmospheric neutrinos.

\mysection{Secondary neutrino sources}
Instead of direct production $\nu_\mu$ may be produced from secondary decays of the DM annihilation products, and we concentrate here on annihilations to charged lepton final states.
Muons are stopped long before they decay~\cite{Ritz:1987mh} and are not a source of high energy neutrinos, whereas taus lose very little energy and will produce prompt neutrinos.  In the case where the DM annihilates preferentially to tau leptons, which in turn decay into neutrinos, the induced muon flux at the Earth surface, taking into account interactions with the material in the Earth, can be parametrized by the following analytic formula \cite{Edsjo:1995wc}:  
\be
\frac{d\Phi_\mu^{\rm S}}{dE_\mu}=B_{\bar{\tau}\tau} \Gamma \frac{p_1m_\chi\, e^{-p_7E_\mu} \left(1-e^{-p_5 m_\chi}\right)}{1+\exp\left[\frac{E_\mu-m\chi\left(p_6+p_2\exp\left(-p_3m_\chi\right)\right)}{p_4m_\chi}\right]}~,
\ee
where $m_\chi$ is in $\gev$.  $B_{\bar{\tau}\tau}$ is the branching ratio of DM annihilating into taus and \\
$p_i\approx(2\times 10^{-22}/\mbox{km}^{2},0.2,5\times 10^{-3},0.1,6\times 10^{-3},0.2,10^{-3})$.

\mysection{Backgrounds}
The main source of background muon neutrinos comes from the shower of cosmic-ray interactions with the atmosphere. The anisotropic induced muon flux is then obtained from \cite{Barger:2007xf,Barger:2007hj}:
\begin{eqnarray}
&&\frac{d^2\Phi_\mu^B}{dE_\mu d\cos\theta_z}=\int_{E_\mu}^\infty dE_\nu\frac{d^2\Phi_\nu}{dE_\nu d\cos\theta_z}R_\mu(E_\mu)R(\cos\theta_z)\nonumber\\
&& \times \left[\frac{d\sigma_\nu^p(E_\nu,E_\mu)}{dE_\mu}\rho_p+ (p\rightarrow n)
\right]+(\nu  \rightarrow \bar{\nu})~,
\end{eqnarray} 
where $\theta_z$ is the zenith angle and the differential fluxes of muon neutrinos and antineutrinos are estimated from the tables found in Honda et al.~\cite{Honda:2006qj}. The function $R(\cos\theta_z)=0.70-0.48\cos\theta_z$ for $\theta_z>85^\circ$ and $1$ elsewhere, is the efficiency of IceCube for tracking up-going muons. The background 
can be substantially reduced by noting that the 
signal 
is collimated in a cone of half-angle, 
$
\Delta\theta = 1.8^\circ(\tev/ E_\nu)^{1/2},
$
about $\theta_z=180^\circ$, where $E_\nu$ is the energy of the incoming neutrino~\cite{Cirelli:2005gh,Jungman:1995df}.  
For the Sun the background reduction is limited by the angular resolution of IceCube, $\Delta\theta=0.5^\circ$, which we take about $\theta_z\simeq 66^\circ$. 
While the primary neutrino signal is monochromatic, $E_\nu=m_\chi$, the spectrum of secondary neutrinos is concentrated at low energy due to the hard slowdown of the DM annihilation products before they decay into neutrinos. Their typical energy is $E_\nu\sim E_{\rm th}=50$ GeV, and $\Delta\theta\simeq 8^\circ$ which increases the relevant background by an order of magnitude compared to the primary neutrino case. Hence, the monochromatic neutrinos offer the best hope for a discovery at IceCube.
\begin{figure}[tp]
\begin{center}
\includegraphics[width=200pt]{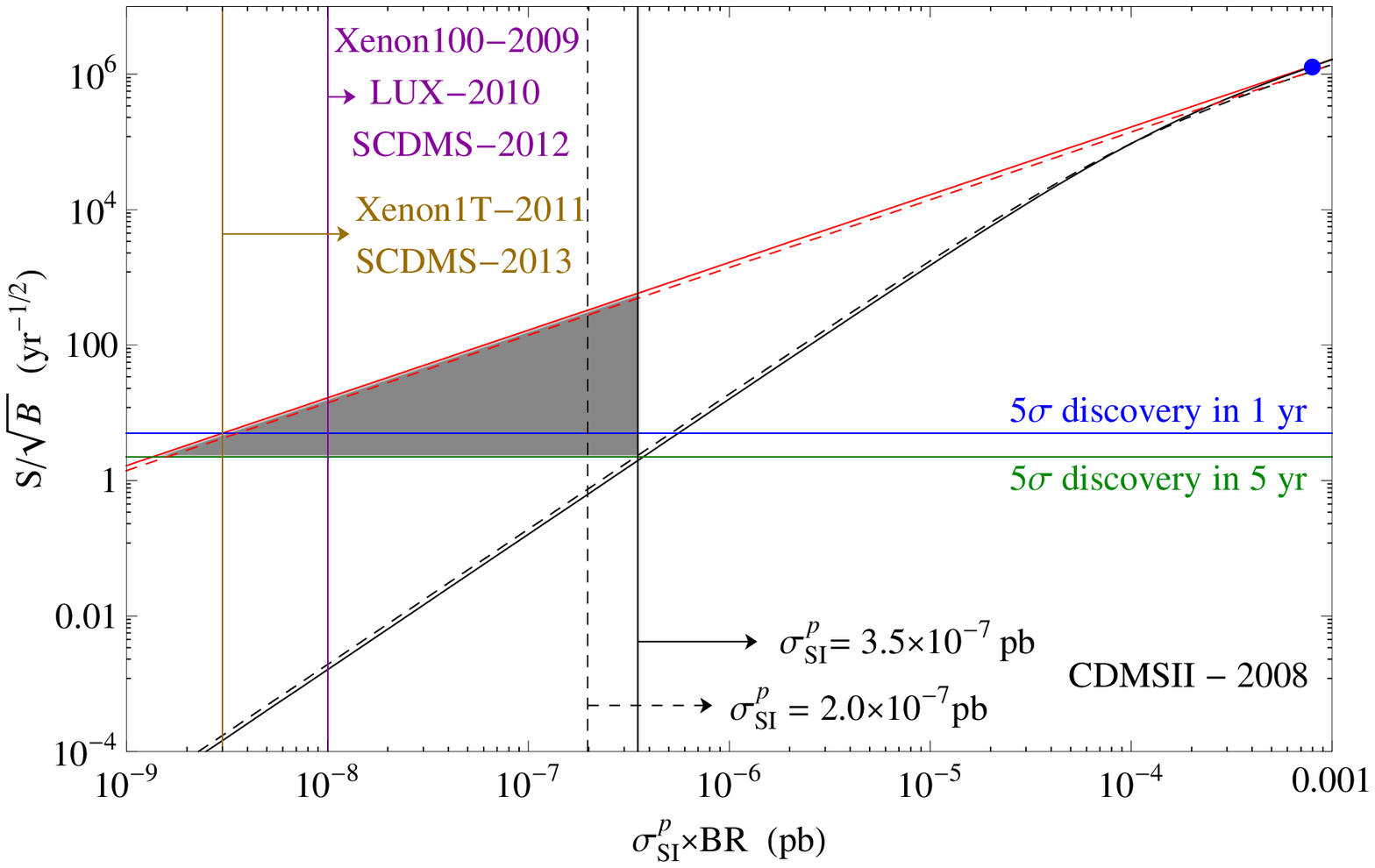}\vskip -0.2cm
\includegraphics[width=200pt]{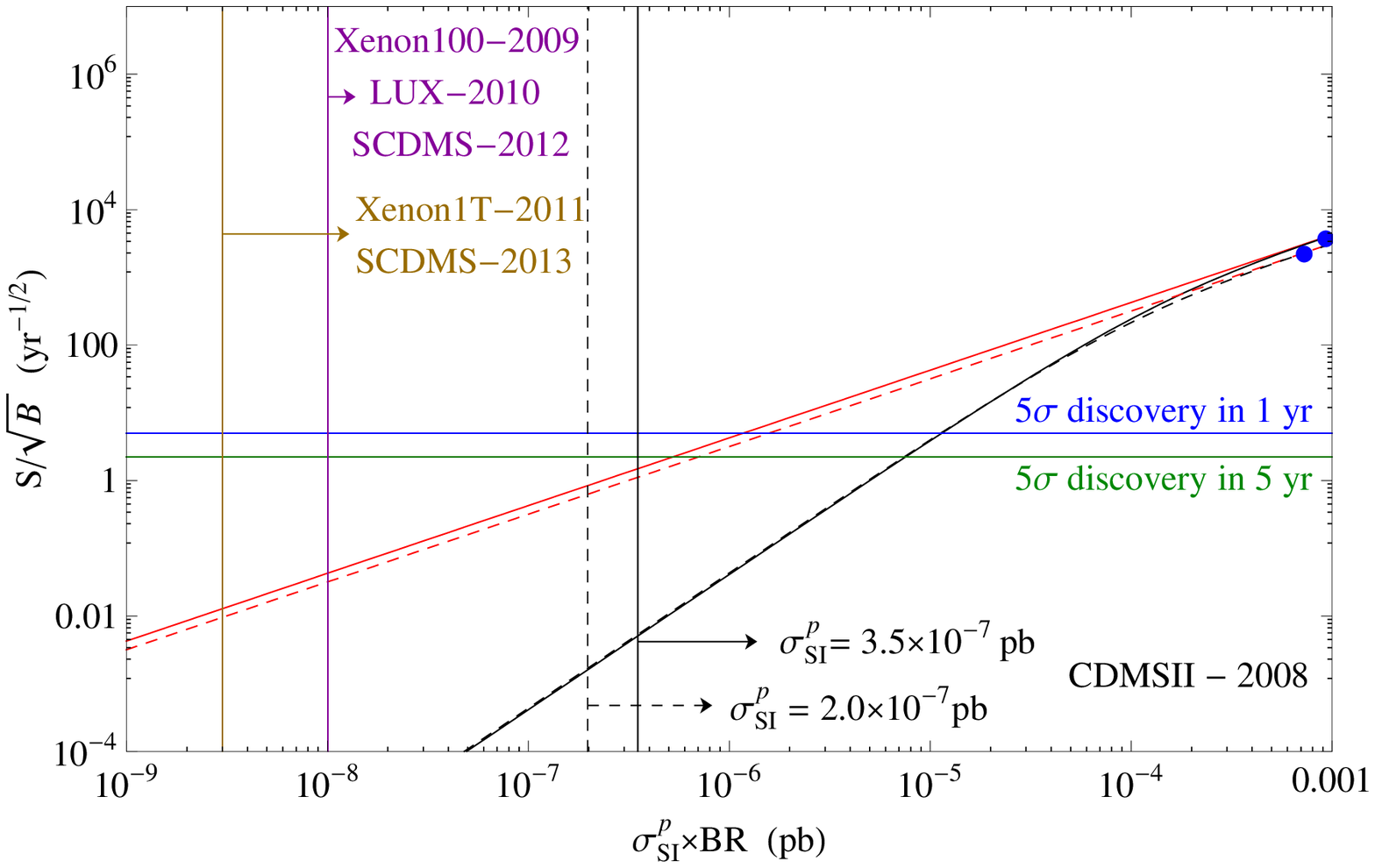}
\caption{Statistical significance of a primary (top) and secondary (bottom) neutrino signals above the atmospheric background.  
BR denotes 
$B_{\bar{\nu}\nu}$ (primary) or $B_{\bar{\tau}\tau}$ (secondary). Upper curves (red) are the equilibrium fluxes 
while lower ones (black) are the naive fluxes. 
The blue dot corresponds to the critical capture rate, $C^{\rm c}_\oplus$, see Eq. (\ref{CcriticEarth}).
The vertical lines show the present and future upper bounds on $\sigma_{\rm SI}^p$ from direct detection. 
The horizontal lines show the discovery reach of IceCube.
}\label{fig:earthsignificance}
\end{center}
\end{figure}

\mysection{Earth and Sun detection potential}
The reach of IceCube is shown in Figure~\ref{fig:earthsignificance} for both primary and secondary neutrinos.
We apply energy cuts for primary (secondary) signals of $250\,\gev<E_\mu<m_\chi$ ($E_{\rm th}<E_\mu< 500\,\gev$). From these plots it is clear that DM that does not annihilate directly to neutrinos has very little hope of discovery at IceCube, even with a large Sommerfeld enhancement, and we concentrate on the primary neutrino case. The maximum neutrino flux is given by the red line and is well into the 5$\sigma$ discovery region for most of the range that can be probed by direct detection. 
However, since $t_\oplus\ll t_{\rm eq}$ the expected rate is denoted by the black line. Enhancements of $\gtap 100$ are necessary for the ATIC/PAMELA results and may, depending on the details of the resonance structure~\cite{Cirelli:2007xd,MarchRussell:2008tu}, be considerably larger for DM in the Earth.  Over most of the region accessible to direct detection enhancement factors of 100-1000 will move us back into the IceCube discovery region.
Thus, by 2013 we will have probed most of the region where neutrinos from the Earth could be discovered.  If direct detection experiments make an observation then we may have a correlated discovery in IceCube.  

\begin{figure}
\begin{center}
\includegraphics[width=200pt]{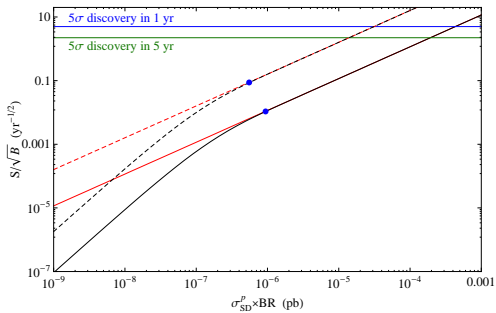}
\caption{Statistical significance for primary neutrino from the Sun as a function of SD scattering off proton. Since the capture rate in the Sun is more efficient than in the Earth it is most likely in equilibrium.}
\label{fig:sunSsqrtB}
\end{center}
\end{figure}

In the core of the Sun the capture rate is dominated by the spin dependent (SD) elastic scattering of DM off hydrogen nuclei~\cite{Jungman:1995df}: 
\bea
C_\odot\simeq  3.57\times 10^{18}\mbox{s}^{-1} 
\frac{\rho^\chi_{0.3}}{\left(v^{\chi}_{270}\right)^3} \left(\frac{\mbox{TeV}}{m_\chi}\right)^2\left(\frac{\sigma^p_{\rm SD}}{10^{-6}\mbox{pb}}\right),
\eea
and~\footnote{We use $V_{eff}= 1.8\times 10^{26} \mathrm{cm}^3 \left(\tev /m_\chi\right)^{3/2}$ for the effective volume of the core of the Sun~\cite{Gould:1987ir}.}
$
C^{\rm c}_\odot\simeq 3.23 \times 10^{17}\mathrm{s}^{-1}\left(\tev/m_\chi\right)^{3/2}.
$
We emphasize again that there are significant astrophysics uncertainties on the DM density and its velocity distributions~\cite{Cirelli:2005gh}, and thus the actual capture rate 
(we use the value from~\cite{Jungman:1995df} for concreteness; see also~\cite{Liu:2008kz}).
Furthermore, the SD scattering cross section of DM on proton is less constrained experimentally, the present bound being $\sigma_{\rm SD}^p\lesssim 0.8$pb  (for $m_\chi\sim 1\,\tev$) from KIMS~\cite{Lee.:2007qn}. Thus, in this case, a future signal from the sun, only from IceCube, would be harder to cleanly interpret. 
It is expected, generically, that the SD cross section is 3-4 order of magnitude larger than the SI one. 
Thus, taking the above capture rates at face value we see that $C^{\rm c}_\odot\ll C_\odot$ for wide range of reasonable DM models.
Consequently, it is likely that the sun is now in equilibrium and its neutrino flux is already maximal, leaving no room for an enhancement of the annihilation rate as shown on 
Figure~\ref{fig:sunSsqrtB}. 
%
%

\mysection{Conclusions}
Combining the information on the neutrino flux and the direct detection cross section
yields a fairly robust measurement of the annihilation boost factor.  The significance of the signal is greatly improved in cases where the annihilation channels involve
primary neutrinos, not an inconvincible possibility\cite{Fox:2008kb}.
A detailed study of a possible signal from primary and secondary Sun-born neutrinos may help to determine the primary branching ratio. In this case, lack of a related earthborn signal would indicate that a low velocity enhancement of the annihilation cross section is not the explanation for the ATIC/PAMELA excess. Instead, one would look for an astrophysics explanation. 
In the ideal case where enough events are observed at IceCube a differential energy information 
could be extracted which may yield further insight towards the DM sector, such as
its mass and decaying branching ratios.

\mysection{Acknowledgements}
We thank D. Hooper, Y. Nir, G. Shaughnessy, T. Volansky, E. Waxman and I. Yavin for  discussions.  Fermilab is operated by Fermi Research Alliance, LLC, under Contract DE-AC02-07CH11359 with the US 
DOE. 

\bibliographystyle{apsrev}
\bibliography{nuearth}

\begin{thebibliography}{47}
\expandafter\ifx\csname natexlab\endcsname\relax\def\natexlab#1{#1}\fi
\expandafter\ifx\csname bibnamefont\endcsname\relax
  \def\bibnamefont#1{#1}\fi
\expandafter\ifx\csname bibfnamefont\endcsname\relax
  \def\bibfnamefont#1{#1}\fi
\expandafter\ifx\csname citenamefont\endcsname\relax
  \def\citenamefont#1{#1}\fi
\expandafter\ifx\csname url\endcsname\relax
  \def\url#1{\texttt{#1}}\fi
\expandafter\ifx\csname urlprefix\endcsname\relax\def\urlprefix{URL }\fi
\providecommand{\bibinfo}[2]{#2}
\providecommand{\eprint}[2][]{\url{#2}}

\bibitem[{\citenamefont{Hooper et~al.}(2008)\citenamefont{Hooper, Blasi, and
  Serpico}}]{Hooper:2008kg}
\bibinfo{author}{\bibfnamefont{D.}~\bibnamefont{Hooper}},
  \bibinfo{author}{\bibfnamefont{P.}~\bibnamefont{Blasi}}, \bibnamefont{and}
  \bibinfo{author}{\bibfnamefont{P.~D.} \bibnamefont{Serpico}}
  (\bibinfo{year}{2008}), \eprint{0810.1527}.

\bibitem[{\citenamefont{Chang et~al.}(2008)}]{:2008zz}
\bibinfo{author}{\bibfnamefont{J.}~\bibnamefont{Chang}} \bibnamefont{et~al.},
  \bibinfo{journal}{Nature} \textbf{\bibinfo{volume}{456}},
  \bibinfo{pages}{362} (\bibinfo{year}{2008}).

\bibitem[{\citenamefont{Adriani et~al.}(2008{\natexlab{a}})}]{Adriani:2008zr}
\bibinfo{author}{\bibfnamefont{O.}~\bibnamefont{Adriani}} \bibnamefont{et~al.}
  (\bibinfo{year}{2008}{\natexlab{a}}), \eprint{0810.4995}.

\bibitem[{\citenamefont{Adriani et~al.}(2008{\natexlab{b}})}]{Adriani:2008zq}
\bibinfo{author}{\bibfnamefont{O.}~\bibnamefont{Adriani}} \bibnamefont{et~al.}
  (\bibinfo{year}{2008}{\natexlab{b}}), \eprint{0810.4994}.

\bibitem[{\citenamefont{Cirelli et~al.}(2007)\citenamefont{Cirelli, Strumia,
  and Tamburini}}]{Cirelli:2007xd}
\bibinfo{author}{\bibfnamefont{M.}~\bibnamefont{Cirelli}},
  \bibinfo{author}{\bibfnamefont{A.}~\bibnamefont{Strumia}}, \bibnamefont{and}
  \bibinfo{author}{\bibfnamefont{M.}~\bibnamefont{Tamburini}},
  \bibinfo{journal}{Nucl. Phys.} \textbf{\bibinfo{volume}{B787}},
  \bibinfo{pages}{152} (\bibinfo{year}{2007}), \eprint{0706.4071}.

\bibitem[{\citenamefont{Cirelli et~al.}(2008)}]{Cirelli:2008pk}
\bibinfo{author}{\bibfnamefont{M.}~\bibnamefont{Cirelli}} \bibnamefont{et~al.}
  (\bibinfo{year}{2008}), \eprint{0809.2409}.

\bibitem[{\citenamefont{Arkani-Hamed et~al.}(2008)}]{ArkaniHamed:2008qn}
\bibinfo{author}{\bibfnamefont{N.}~\bibnamefont{Arkani-Hamed}}
  \bibnamefont{et~al.} (\bibinfo{year}{2008}), \eprint{0810.0713}.

\bibitem[{\citenamefont{Cirelli and Strumia}(2008)}]{Cirelli:2008jk}
\bibinfo{author}{\bibfnamefont{M.}~\bibnamefont{Cirelli}} \bibnamefont{and}
  \bibinfo{author}{\bibfnamefont{A.}~\bibnamefont{Strumia}}
  (\bibinfo{year}{2008}), \eprint{0808.3867}.

\bibitem[{\citenamefont{Cholis et~al.}(2008)}]{Cholis:2008hb}
\bibinfo{author}{\bibfnamefont{I.}~\bibnamefont{Cholis}} \bibnamefont{et~al.}
  (\bibinfo{year}{2008}), \eprint{0809.1683}.

\bibitem[{\citenamefont{Feldman et~al.}(2008)\citenamefont{Feldman, Liu, and
  Nath}}]{Feldman:2008xs}
\bibinfo{author}{\bibfnamefont{D.}~\bibnamefont{Feldman}},
  \bibinfo{author}{\bibfnamefont{Z.}~\bibnamefont{Liu}}, \bibnamefont{and}
  \bibinfo{author}{\bibfnamefont{P.}~\bibnamefont{Nath}}
  (\bibinfo{year}{2008}), \eprint{0810.5762}.

\bibitem[{\citenamefont{Ibe et~al.}(2008)\citenamefont{Ibe, Murayama, and
  Yanagida}}]{Ibe:2008ye}
\bibinfo{author}{\bibfnamefont{M.}~\bibnamefont{Ibe}},
  \bibinfo{author}{\bibfnamefont{H.}~\bibnamefont{Murayama}}, \bibnamefont{and}
  \bibinfo{author}{\bibfnamefont{T.~T.} \bibnamefont{Yanagida}}
  (\bibinfo{year}{2008}), \eprint{0812.0072}.

\bibitem[{\citenamefont{Collaboration}(2008)}]{Collaboration:2008aa}
\bibinfo{author}{\bibfnamefont{H.~E. S.~S.} \bibnamefont{Collaboration}}
  (\bibinfo{year}{2008}), \eprint{0811.3894}.

\bibitem[{\citenamefont{Hinton}(2004)}]{Hinton:2004eu}
\bibinfo{author}{\bibfnamefont{J.~A.} \bibnamefont{Hinton}}
  (\bibinfo{collaboration}{The HESS}), \bibinfo{journal}{New Astron. Rev.}
  \textbf{\bibinfo{volume}{48}}, \bibinfo{pages}{331} (\bibinfo{year}{2004}),
  \eprint{astro-ph/0403052}.

\bibitem[{\citenamefont{de~Boer et~al.}(2005)}]{deBoer:2005tm}
\bibinfo{author}{\bibfnamefont{W.}~\bibnamefont{de~Boer}} \bibnamefont{et~al.},
  \bibinfo{journal}{Astron. Astrophys.} \textbf{\bibinfo{volume}{444}},
  \bibinfo{pages}{51} (\bibinfo{year}{2005}), \eprint{astro-ph/0508617}.

\bibitem[{\citenamefont{Elsaesser and Mannheim}(2005)}]{Elsaesser:2004xa}
\bibinfo{author}{\bibfnamefont{D.}~\bibnamefont{Elsaesser}} \bibnamefont{and}
  \bibinfo{author}{\bibfnamefont{K.}~\bibnamefont{Mannheim}}
  (\bibinfo{collaboration}{MAGIC}), \bibinfo{journal}{New Astron. Rev.}
  \textbf{\bibinfo{volume}{49}}, \bibinfo{pages}{297} (\bibinfo{year}{2005}),
  \eprint{astro-ph/0409563}.

\bibitem[{\citenamefont{Cecchi}(2008)}]{Cecchi:2008zz}
\bibinfo{author}{\bibfnamefont{C.}~\bibnamefont{Cecchi}}
  (\bibinfo{collaboration}{GLAST LAT}), \bibinfo{journal}{J. Phys. Conf. Ser.}
  \textbf{\bibinfo{volume}{120}}, \bibinfo{pages}{062017}
  (\bibinfo{year}{2008}).

\bibitem[{\citenamefont{Boezio et~al.}(2001)}]{Boezio:2001ac}
\bibinfo{author}{\bibfnamefont{M.}~\bibnamefont{Boezio}} \bibnamefont{et~al.}
  (\bibinfo{collaboration}{WiZard/CAPRICE}), \bibinfo{journal}{Astrophys. J.}
  \textbf{\bibinfo{volume}{561}}, \bibinfo{pages}{787} (\bibinfo{year}{2001}),
  \eprint{astro-ph/0103513}.

\bibitem[{\citenamefont{Barwick et~al.}(1997)}]{Barwick:1997ig}
\bibinfo{author}{\bibfnamefont{S.~W.} \bibnamefont{Barwick}}
  \bibnamefont{et~al.} (\bibinfo{collaboration}{HEAT}),
  \bibinfo{journal}{Astrophys. J.} \textbf{\bibinfo{volume}{482}},
  \bibinfo{pages}{L191} (\bibinfo{year}{1997}), \eprint{astro-ph/9703192}.

\bibitem[{\citenamefont{Aguilar et~al.}(2007)}]{Aguilar:2007yf}
\bibinfo{author}{\bibfnamefont{M.}~\bibnamefont{Aguilar}} \bibnamefont{et~al.}
  (\bibinfo{collaboration}{AMS-01}), \bibinfo{journal}{Phys. Lett.}
  \textbf{\bibinfo{volume}{B646}}, \bibinfo{pages}{145} (\bibinfo{year}{2007}),
  \eprint{astro-ph/0703154}.

\bibitem[{\citenamefont{Liu et~al.}(2008{\natexlab{a}})\citenamefont{Liu, Yin,
  and Zhu}}]{Liu:2008ci}
\bibinfo{author}{\bibfnamefont{J.}~\bibnamefont{Liu}},
  \bibinfo{author}{\bibfnamefont{P.-f.} \bibnamefont{Yin}}, \bibnamefont{and}
  \bibinfo{author}{\bibfnamefont{S.-h.} \bibnamefont{Zhu}}
  (\bibinfo{year}{2008}{\natexlab{a}}), \eprint{0812.0964}.

\bibitem[{\citenamefont{Hisano et~al.}(2008)}]{Hisano:2008ah}
\bibinfo{author}{\bibfnamefont{J.}~\bibnamefont{Hisano}} \bibnamefont{et~al.}
  (\bibinfo{year}{2008}), \eprint{0812.0219}.

\bibitem[{\citenamefont{Gould}(1987)}]{Gould:1987ir}
\bibinfo{author}{\bibfnamefont{A.}~\bibnamefont{Gould}},
  \bibinfo{journal}{Astrophys. J.} \textbf{\bibinfo{volume}{321}},
  \bibinfo{pages}{571} (\bibinfo{year}{1987}).

\bibitem[{\citenamefont{Lundberg and Edsjo}(2004)}]{Lundberg:2004dn}
\bibinfo{author}{\bibfnamefont{J.}~\bibnamefont{Lundberg}} \bibnamefont{and}
  \bibinfo{author}{\bibfnamefont{J.}~\bibnamefont{Edsjo}},
  \bibinfo{journal}{Phys. Rev.} \textbf{\bibinfo{volume}{D69}},
  \bibinfo{pages}{123505} (\bibinfo{year}{2004}), \eprint{astro-ph/0401113}.

\bibitem[{\citenamefont{Achterberg et~al.}(2006)}]{Achterberg:2006md}
\bibinfo{author}{\bibfnamefont{A.}~\bibnamefont{Achterberg}}
  \bibnamefont{et~al.} (\bibinfo{collaboration}{IceCube}),
  \bibinfo{journal}{Astropart. Phys.} \textbf{\bibinfo{volume}{26}},
  \bibinfo{pages}{155} (\bibinfo{year}{2006}), \eprint{astro-ph/0604450}.

\bibitem[{\citenamefont{Ahmed et~al.}(2008)}]{Ahmed:2008eu}
\bibinfo{author}{\bibfnamefont{Z.}~\bibnamefont{Ahmed}} \bibnamefont{et~al.}
  (\bibinfo{collaboration}{CDMS}) (\bibinfo{year}{2008}), \eprint{0802.3530}.

\bibitem[{\citenamefont{Angle et~al.}(2008)}]{Angle:2007uj}
\bibinfo{author}{\bibfnamefont{J.}~\bibnamefont{Angle}} \bibnamefont{et~al.}
  (\bibinfo{collaboration}{XENON}), \bibinfo{journal}{Phys. Rev. Lett.}
  \textbf{\bibinfo{volume}{100}}, \bibinfo{pages}{021303}
  (\bibinfo{year}{2008}), \eprint{0706.0039}.

\bibitem[{\citenamefont{Jungman et~al.}(1996)\citenamefont{Jungman,
  Kamionkowski, and Griest}}]{Jungman:1995df}
\bibinfo{author}{\bibfnamefont{G.}~\bibnamefont{Jungman}},
  \bibinfo{author}{\bibfnamefont{M.}~\bibnamefont{Kamionkowski}},
  \bibnamefont{and} \bibinfo{author}{\bibfnamefont{K.}~\bibnamefont{Griest}},
  \bibinfo{journal}{Phys. Rept.} \textbf{\bibinfo{volume}{267}},
  \bibinfo{pages}{195} (\bibinfo{year}{1996}), \eprint{hep-ph/9506380}.

\bibitem[{\citenamefont{Tucker-Smith and Weiner}(2001)}]{TuckerSmith:2001hy}
\bibinfo{author}{\bibfnamefont{D.}~\bibnamefont{Tucker-Smith}}
  \bibnamefont{and} \bibinfo{author}{\bibfnamefont{N.}~\bibnamefont{Weiner}},
  \bibinfo{journal}{Phys. Rev.} \textbf{\bibinfo{volume}{D64}},
  \bibinfo{pages}{043502} (\bibinfo{year}{2001}), \eprint{hep-ph/0101138}.

\bibitem[{\citenamefont{Peter and Tremaine}(2008)}]{Peter:2008sy}
\bibinfo{author}{\bibfnamefont{A.~H.~G.} \bibnamefont{Peter}} \bibnamefont{and}
  \bibinfo{author}{\bibfnamefont{S.}~\bibnamefont{Tremaine}}
  (\bibinfo{year}{2008}), \eprint{0806.2133}.

\bibitem[{\citenamefont{Damour and Krauss}(1999)}]{Damour:1998vg}
\bibinfo{author}{\bibfnamefont{T.}~\bibnamefont{Damour}} \bibnamefont{and}
  \bibinfo{author}{\bibfnamefont{L.~M.} \bibnamefont{Krauss}},
  \bibinfo{journal}{Phys. Rev.} \textbf{\bibinfo{volume}{D59}},
  \bibinfo{pages}{063509} (\bibinfo{year}{1999}), \eprint{astro-ph/9807099}.

\bibitem[{\citenamefont{Xu and Siegel}(2008)}]{Xu:2008ep}
\bibinfo{author}{\bibfnamefont{X.}~\bibnamefont{Xu}} \bibnamefont{and}
  \bibinfo{author}{\bibfnamefont{E.~R.} \bibnamefont{Siegel}}
  (\bibinfo{year}{2008}), \eprint{0806.3767}.

\bibitem[{\citenamefont{Vogelsberger et~al.}(2008)}]{Vogelsberger:2008qb}
\bibinfo{author}{\bibfnamefont{M.}~\bibnamefont{Vogelsberger}}
  \bibnamefont{et~al.} (\bibinfo{year}{2008}), \eprint{0812.0362}.

\bibitem[{\citenamefont{Brink et~al.}(2005)}]{Brink:2005ej}
\bibinfo{author}{\bibfnamefont{P.~L.} \bibnamefont{Brink}} \bibnamefont{et~al.}
  (\bibinfo{collaboration}{CDMS-II}) (\bibinfo{year}{2005}),
  \eprint{astro-ph/0503583}.

\bibitem[{\citenamefont{Aprile et~al.}(2002)}]{Aprile:2002ef}
\bibinfo{author}{\bibfnamefont{E.}~\bibnamefont{Aprile}} \bibnamefont{et~al.}
  (\bibinfo{year}{2002}), \eprint{astro-ph/0207670}.

\bibitem[{\citenamefont{http://lux.brown.edu/index.html}(2008)}]{LUX}
\bibinfo{author}{\bibnamefont{http://lux.brown.edu/index.html}}
  (\bibinfo{year}{2008}).

\bibitem[{\citenamefont{Barger et~al.}(2007)}]{Barger:2007xf}
\bibinfo{author}{\bibfnamefont{V.}~\bibnamefont{Barger}} \bibnamefont{et~al.},
  \bibinfo{journal}{Phys. Rev.} \textbf{\bibinfo{volume}{D76}},
  \bibinfo{pages}{095008} (\bibinfo{year}{2007}), \eprint{0708.1325}.

\bibitem[{\citenamefont{Gonzalez-Garcia
  et~al.}(2005)\citenamefont{Gonzalez-Garcia, Halzen, and
  Maltoni}}]{GonzalezGarcia:2005xw}
\bibinfo{author}{\bibfnamefont{M.~C.} \bibnamefont{Gonzalez-Garcia}},
  \bibinfo{author}{\bibfnamefont{F.}~\bibnamefont{Halzen}}, \bibnamefont{and}
  \bibinfo{author}{\bibfnamefont{M.}~\bibnamefont{Maltoni}},
  \bibinfo{journal}{Phys. Rev.} \textbf{\bibinfo{volume}{D71}},
  \bibinfo{pages}{093010} (\bibinfo{year}{2005}), \eprint{hep-ph/0502223}.

\bibitem[{\citenamefont{Ritz and Seckel}(1988)}]{Ritz:1987mh}
\bibinfo{author}{\bibfnamefont{S.}~\bibnamefont{Ritz}} \bibnamefont{and}
  \bibinfo{author}{\bibfnamefont{D.}~\bibnamefont{Seckel}},
  \bibinfo{journal}{Nucl. Phys.} \textbf{\bibinfo{volume}{B304}},
  \bibinfo{pages}{877} (\bibinfo{year}{1988}).

\bibitem[{\citenamefont{Edsjo}(1995)}]{Edsjo:1995wc}
\bibinfo{author}{\bibfnamefont{J.}~\bibnamefont{Edsjo}},
  \bibinfo{journal}{Nucl. Phys. Proc. Suppl.} \textbf{\bibinfo{volume}{43}},
  \bibinfo{pages}{265} (\bibinfo{year}{1995}), \eprint{hep-ph/9504205}.

\bibitem[{\citenamefont{Barger et~al.}(2008)\citenamefont{Barger, Keung, and
  Shaughnessy}}]{Barger:2007hj}
\bibinfo{author}{\bibfnamefont{V.~D.} \bibnamefont{Barger}},
  \bibinfo{author}{\bibfnamefont{W.-Y.} \bibnamefont{Keung}}, \bibnamefont{and}
  \bibinfo{author}{\bibfnamefont{G.}~\bibnamefont{Shaughnessy}},
  \bibinfo{journal}{Phys. Lett.} \textbf{\bibinfo{volume}{B664}},
  \bibinfo{pages}{190} (\bibinfo{year}{2008}), \eprint{0709.3301}.

\bibitem[{\citenamefont{Honda et~al.}(2007)}]{Honda:2006qj}
\bibinfo{author}{\bibfnamefont{M.}~\bibnamefont{Honda}} \bibnamefont{et~al.},
  \bibinfo{journal}{Phys. Rev.} \textbf{\bibinfo{volume}{D75}},
  \bibinfo{pages}{043006} (\bibinfo{year}{2007}), \eprint{astro-ph/0611418}.

\bibitem[{\citenamefont{Cirelli et~al.}(2005)}]{Cirelli:2005gh}
\bibinfo{author}{\bibfnamefont{M.}~\bibnamefont{Cirelli}} \bibnamefont{et~al.},
  \bibinfo{journal}{Nucl. Phys.} \textbf{\bibinfo{volume}{B727}},
  \bibinfo{pages}{99} (\bibinfo{year}{2005}), \eprint{hep-ph/0506298}.

\bibitem[{\citenamefont{March-Russell and West}(2008)}]{MarchRussell:2008tu}
\bibinfo{author}{\bibfnamefont{J.~D.} \bibnamefont{March-Russell}}
  \bibnamefont{and} \bibinfo{author}{\bibfnamefont{S.~M.} \bibnamefont{West}}
  (\bibinfo{year}{2008}), \eprint{0812.0559}.

\bibitem[{\citenamefont{Liu et~al.}(2008{\natexlab{b}})\citenamefont{Liu, Yin,
  and Zhu}}]{Liu:2008kz}
\bibinfo{author}{\bibfnamefont{J.}~\bibnamefont{Liu}},
  \bibinfo{author}{\bibfnamefont{P.-f.} \bibnamefont{Yin}}, \bibnamefont{and}
  \bibinfo{author}{\bibfnamefont{S.-h.} \bibnamefont{Zhu}},
  \bibinfo{journal}{Phys. Rev.} \textbf{\bibinfo{volume}{D77}},
  \bibinfo{pages}{115014} (\bibinfo{year}{2008}{\natexlab{b}}),
  \eprint{0803.2164}.

\bibitem[{\citenamefont{Lee. et~al.}(2007)}]{Lee.:2007qn}
\bibinfo{author}{\bibfnamefont{H.~S.} \bibnamefont{Lee.}} \bibnamefont{et~al.}
  (\bibinfo{collaboration}{KIMS}), \bibinfo{journal}{Phys. Rev. Lett.}
  \textbf{\bibinfo{volume}{99}}, \bibinfo{pages}{091301}
  (\bibinfo{year}{2007}), \eprint{0704.0423}.

\bibitem[{\citenamefont{Fox and Poppitz}(2008)}]{Fox:2008kb}
\bibinfo{author}{\bibfnamefont{P.~J.} \bibnamefont{Fox}} \bibnamefont{and}
  \bibinfo{author}{\bibfnamefont{E.}~\bibnamefont{Poppitz}}
  (\bibinfo{year}{2008}), \eprint{0811.0399}.

\bibitem[{\citenamefont{Dutta et~al.}(2001)}]{Dutta:2000hh}
\bibinfo{author}{\bibfnamefont{S.~I.} \bibnamefont{Dutta}}
  \bibnamefont{et~al.}, \bibinfo{journal}{Phys. Rev.}
  \textbf{\bibinfo{volume}{D63}}, \bibinfo{pages}{094020}
  (\bibinfo{year}{2001}), \eprint{hep-ph/0012350}.

\end{thebibliography}

\end{document}